\documentclass[11pt,a4paper,twoside]{article}
\usepackage{mathpazo}
\usepackage[a4paper,textwidth=13.5cm,top=3cm,bottom=3.2cm,twoside=false]{geometry}
\usepackage{xcolor}\definecolor{accent}{HTML}{000000}
\usepackage[utf8]{inputenc}
\usepackage[T1]{fontenc}
\usepackage[english]{babel}
\usepackage{amsmath,amssymb}
\usepackage{graphicx,booktabs}
\usepackage{tikz}\usetikzlibrary{arrows.meta}
\usepackage[font=small,labelfont=bf,labelsep=space]{caption}
\usepackage{csquotes}
\usepackage[hang,flushmargin]{footmisc}
\usepackage{microtype}
\usepackage{xcolor}
\usepackage{xurl}
\usepackage[colorlinks,allcolors=accent,hyperfootnotes=false]{hyperref}
\newenvironment{refs}{\section*{References}\small\begin{list}{}{\leftmargin1.5em\itemindent-1.5em\itemsep2pt\parsep0pt}\let\bibitem\item}{\end{list}}

\usepackage{titlesec}
\titleformat{\section}{\normalfont\large\scshape\color{accent}}{\thesection}{1em}{\MakeLowercase}
\titlespacing*{\section}{0pt}{2.2ex plus .5ex}{1.2ex}
\titleformat{\subsection}{\normalfont\itshape}{\thesubsection}{1em}{}
\usepackage{fancyhdr}

\begin{document}
\thispagestyle{empty}
\begin{center}
\vspace*{1em}
{\large\bfseries Schrödinger's Legacy: The Wave Function and Quantum Reality\par}
\vspace{1.4em}
{\scshape Davide Romano{\renewcommand{\thefootnote}{\fnsymbol{footnote}}\footnotemark[1]} \enspace{\normalfont$\cdot$}\enspace Massimiliano Badino}%
{\renewcommand{\thefootnote}{\fnsymbol{footnote}}\footnote{University of Verona, Department of Human Sciences, Lungadige Porta Vittoria 17, 37129 Verona.\newline Email: \href{mailto:davide.romano@univr.it}{\nolinkurl{davide.romano@univr.it}}, \href{mailto:massimiliano.badino@univr.it}{\nolinkurl{massimiliano.badino@univr.it}}}\addtocounter{footnote}{-1}}\par
\vspace{1em}
{\color{accent}\rule{3cm}{0.4pt}}\par\vspace{0.6em}
{\small Forthcoming in R.~Arroyo \& M.~Morganti (eds.), \emph{Quantum Dilemmas:\\ Philosophical Issues in Quantum Theory after One Century}, Springer.\par}
\end{center}
\vspace{1.2em}
\begin{quote}\noindent{\scshape\color{accent} Abstract.}\enspace The article discusses how Schrödinger arrived at the formulation of wave
mechanics and his legacy on the philosophy of quantum mechanics,
focusing in particular on the metaphysics of the wave function.
Schrödinger's interpretation of the wave function as electron charge
density is generally dismissed or presented only as a curious historical
example in the literature. On a closer look, however, while this
interpretation was eventually abandoned in favor of Born's statistical
interpretation, Schrödinger's idea survived in modern realist approaches
to the wave function, such as the mass-density approach and the
multi-field interpretation. It also proved to be of great utility in
modern techniques and approximations in quantum chemistry to compute the
shape of atoms and molecules. Overall, Schrödinger's interpretation of
the wave function and his realist understanding of quantum mechanics
paved the way to modern attempts to define a clear ontology of the
microscopic world, a tradition that survived the Copenhagen
interpretation of Bohr and Heisenberg and that is still present today.\end{quote}
\vspace{0.8em}
\hypertarget{introduction}{%
\section{Introduction}}

Erwin Schrödinger is one of the founding fathers of quantum mechanics.
From his early works on quantum theory, he attempted to find an
alternative formulation with respect to the matrix mechanics developed
by Heisenberg, Born and Jordan in 1925.\footnote{Heisenberg (1925); Born and Jordan (1925); Born, Heisenberg and
  Jordan (1926).} Schrödinger's motivation for
such an attempt was to search for a theory in continuity with classical
mechanics, a theory describing continuous processes in space and that
did not force us to abandon entirely the ontology of the microscopic
world. He developed such an alternative formulation, known as \emph{wave
mechanics}, in four papers in 1926. In these works, he derived the
central dynamical equation of quantum mechanics, i.e.~the \emph{Schrödinger
equation}, and the \emph{wave function} as the mathematical representation of
a quantum system. Right from the beginning, Schrödinger realized that
the newly developed wave mechanics was not as intuitive as he sought,
certainly not as intuitive as wave optics or classical mechanics. The
main difference with previous classical theories was the nature of the
wave function: this new entity was mathematically defined in the
system's configuration space and, for a many-particle system (described
by a \(n\)-particle entangled state), could not be
factorized without losing information on the system state. In other
words, even if wave mechanics looked like a theory of real waves, there
were no waves in three-dimensional space, or at least no classical
waves. The wave function was something new in the ontology of physics.
Still, Schrödinger believed that the wave function had to be part of the
ontology of physics, or (in more modern terms) he was a realist on the
wave function. To this end, he proposed to regard the absolute square of
the wave function multiplied by the charge of the electron\footnote{Schrödinger considers the charge as a property and the electron as
  a specific particle. Probably this was just an initial case study to
  see whether the interpretation was feasible. In the letter of June 6
  to Lorentz he explicitly states to consider ``a certain property of
  the particle''. This suggests he had in mind to generalize this
  definition.}:
\begin{equation}
-e|\psi|^2 \label{eq:charge}
\end{equation}
as the \emph{electron charge density} or equivalently the \emph{density of
electricity}. Both for one particle and many particle systems, the
electron charge density was defined in three-dimensional space, and
taken to be the ontology associated with the wave function (the latter
being just a formal tool to compute the real charge density). This kind
of interpretation was abandoned however in favor of the statistical
interpretation proposed by Born, which became part of the postulates of
quantum mechanics (intertwined with the collapse postulate).
Nevertheless, the ``battle for the ontology'' was not lost, or not
entirely: Schrödinger's idea to connect the wave function to a physical
entity in three-dimensional space survived and in recent times has been
used in various occasions, such as the notion of mass-density introduced
by Ghirardi et al.~(1995) in the spontaneous collapse models or the
multi-field interpretation in Bohm's theory (Forrest 1988; Belot 2012;
Hubert \& Romano 2018).

The article is divided into two parts: the first part (section 2)
discusses the historical context and the principal achievements of early
quantum mechanics (sect. 2.1) which prepares the route to the
formulation of wave mechanics developed by Schrödinger in a series of
four articles published in 1926 (sect. 2.2). The second part (sections
3, 4, 5) discusses Schrödinger's interpretation of the wave function and
the early Lorentz objections (sect. 3); the importance of such an
interpretation for techniques and approximations in quantum chemistry
(sect. 4) and the ``red line'' connecting Schrödinger's charge density
interpretation with modern realist interpretations of the wave function
(sect. 5).

\hypertarget{the-road-to-wave-mechanics}{%
\section{The Road to Wave Mechanics}}

\hypertarget{preparing-wave-mechanics}{%
\subsection{Preparing Wave Mechanics}}

The analysis of Schrödinger's route to wave mechanics has long
constituted a central and enduring problem for historians of modern
physics. From the late 1970s onward, scholars have undertaken extensive
efforts to assemble and critically examine both published writings and
archival materials---correspondence, notebooks, draft manuscripts---to
reconstruct the conceptual path that culminated in one of the most
unexpected and transformative theoretical achievements of
twentieth-century science.\footnote{Within the huge literature, we will especially refer to Hanle
  (1977), Wessels (1979), Mehra and Rechenberg (1987a, 1987b), Joas
  and Lehner (2009) and Duncan and Janssen (2023).} While important interpretive
disagreements persist, and certain documentary lacunae remain, a broad
consensus has gradually emerged regarding the principal stages and
intellectual influences that shaped Schrödinger's breakthrough.

In this introductory discussion, we distinguish between two analytically
separable yet historically intertwined moments in this development.
First, we examine the constellation of sources---conceptual, technical,
and philosophical---that arguably guided Schrödinger toward formulating
a wave-based approach to quantum phenomena. These include the analysis
of Einstein's theory of monoatomic gas, the impact of de Broglie's
matter-wave hypothesis, and the formal analogy between classical
mechanics and ray optics. In the subsequent sections, we turn from
genesis to interpretation, focusing on the distinctive features of
Schrödinger's understanding of his own formalism.

In 1924, Albert Einstein published a short article that, rather
unexpectedly, solved a stubborn riddle of the old quantum theory\footnote{Einstein (1924); Simon (2006, Doc. 27).}. It
was well known that Planck's quantum relation could be employed to
extend the classical statistical-mechanical treatment of entropy. Yet
the derived extensions proved deeply problematic. In particular, they
were able to satisfy two crucial requirements---namely, the extensivity
of entropy and its vanishing (or well-defined limiting behavior) at very
low temperatures in accordance with Nernst's theorem---only at the price
of introducing ad hoc assumptions. Inspired by Bose's derivation of
Planck's radiation law, Einstein recognized that the crux of the matter
lay in a fundamental revision of the statistical counting of microscopic
states. In Boltzmann's classical framework, a microstate is defined by
specifying the energy of each individual particle, thereby presupposing
their distinguishability. Bose, by contrast, counted only the number of
quanta occupying a given energy level, without regard to their
individual identities. What mattered was not \emph{which} particle had a
certain energy, but \emph{how many} particles shared it. By replacing
Boltzmann's combinatorial procedure with Bose's new mode of counting,
Einstein obtained an expression for the gas entropy that automatically
satisfied both the requirement of extensivity and the low-temperature
behavior prescribed by Nernst's theorem.\footnote{Einstein (1925, pp.~5--7); Simon (2006, Doc. 28).}

Getting the right answer was not enough, though. When Schrödinger read
Einstein's first paper on the monoatomic gas, he was, quite
understandably, taken aback. Formed within the honorable tradition of
classical Viennese physics, he regarded Boltzmann's statistics as the
``natural'' framework for statistical mechanics and tended to view
alternative counting procedures as no more than formal or calculational
devices.\footnote{Scott (1967, chap.~2).} For Schrödinger, statistical methods provided a genuine
window onto microscopic reality, and Boltzmann's combinatorial scheme
carried with it a clear ontological commitment: it presupposed the
existence of individual, distinguishable, and independent particles. By
contrast, the Bose--Einstein counting appeared to relinquish precisely
these features, abandoning both individuality and statistical
independence at the fundamental level. Troubled by these implications,
Schrödinger voiced his objections in a letter to Einstein dated 5
February 1925. Einstein replied on 28 February, arguing in detail the
rationale of the new statistics but ultimately agreeing with Schrödinger
that it required a new ontological underpinning (on this exchange, see
also Acuña, this volume). He pointed Schrödinger to his second paper on
gas theory and, in particular, to the argument developed in section 8,
where he calculated the fluctuations in the number of particles within
an infinitesimal energy interval using the new statistics. The result
was striking: the fluctuation term decomposed into two contributions,
one corpuscular and the other undulatory.\footnote{Einstein (1925, pp.~8--11).} According to Einstein, the
most straightforward way to make sense of this disconcerting outcome was
``by suitably associating a radiation process with the gas,'' a move that
``involves more than a mere analogy'' insofar as it points toward a
radically novel ontology. Einstein, therefore, urged Schrödinger to
study Louis de Broglie's thesis\footnote{De Broglie (1924).} carefully, and Schrödinger complied
diligently in the weeks that followed.

The reading was a veritable breakthrough. De Broglie's arguments were
concise and straightforward.\footnote{For a detailed discussion of de Broglie's theory, see Duncan and
  Janssen (2023, pp.~362--371).} He began with noticing that, for a
microscopic particle moving at relativistic speed, for example, the
electron bound in an atomic orbit, its energy can be written in two
different ways: the usual relativistic value ($E=mc^2$) and the
Planck relation ($E=h\nu$). Since both expressions refer to the same
physical quantity, they can be equated. To de Broglie, this equivalence
implied that a phase wave could always be associated with a moving
particle. In turn, such a phase wave must be understood as a ``wave
packet'' composed of several monochromatic components, whose
superposition gives rise to a localized structure. The group velocity of
this packet can then be naturally identified with the velocity of the
moving particle, thereby providing a coherent dynamical link between
particle motion and wave propagation. In the second chapter of his
thesis, de Broglie offers a clearer foundation for the phase wave
concept by rediscovering Hamilton's optical-mechanical analogy. In
1834--35, William Hamilton recognized a profound connection between
Fermat's principle---according to which a light ray traveling from point
A to point B follows the path of shortest optical length---and
Maupertuis's principle, which states that a material particle moves
along a trajectory that renders the mechanical action stationary.
Building on the formal analogy between these extremal principles,
Hamilton developed an elegant reformulation of classical
mechanics---later refined by Jacobi---in which mechanical trajectories
can be interpreted as vectors normal to surfaces on which Hamilton's
characteristic function is constant. During this analysis, Hamilton also
observed that this function possesses the physical dimensions of action,
thereby strengthening the structural parallel between optics and
mechanics.

In Chapter 2 of his thesis, de Broglie applies Hamilton's argument to
the motion of the phase wave and the electron, finding, unsurprisingly,
the equivalence between the two principles: ``Fermat's principle, when
applied to the phase wave, is identical to Maupertuis's principle
applied to the particle; the dynamically possible trajectories of the
particle are identical to the possible rays of the wave.''\footnote{De Broglie (1924, p.~56).} Armed
with this analytical framework, de Broglie was able to derive two
far-reaching results. First, he provided a novel derivation of the
Bohr--Sommerfeld stability conditions by requiring that an electron
moving along a closed orbit can remain dynamically stable only if the
length of that orbit is ``in resonance'' with its phase wave---that is,
only if the wave fits along the orbit in such a way that an integer
number of wavelengths is completed after one full revolution. Under this
condition, the phase forms a standing wave along the trajectory. The
classical image of a continuously variable motion is thereby replaced
with a discrete set of dynamically admissible orbits, since only certain
resonant configurations are permitted. In this way, de Broglie was able
to recover the quantization rules of the Bohr--Sommerfeld theory from a
unified wave--particle perspective, rather than postulating them
independently.

Second, in Chapter 7 of his thesis, de Broglie extended this line of
reasoning beyond atomic structure to the domains of gas statistics and
blackbody radiation. Here, the guiding idea is both simple and profound:
a vessel filled with gas molecules must also be understood as a vessel
permeated by their associated phase waves. The thermodynamic and
statistical behavior of the gas can therefore no longer be analyzed
purely in corpuscular terms. Instead, one must take into account the
wave phenomena---interference, resonance, and the formation of
stationary modes---that arise from the superposition of these phase
waves:

\begin{quote}
If the motion of gaseous atoms is accompanied by the propagation of
waves, the vessel containing the gas will be traversed in all
directions by these waves. We are thus naturally led---much as in the
conception of black-body radiation developed by Mr.~Jeans---to regard
the phase waves that form stationary systems (that is, that resonate
with the dimensions of the enclosure) as the only stable ones. These
alone would enter into the study of thermodynamic equilibrium.\footnote{De Broglie (1924, p.~110).}
\end{quote}

It is precisely the intimate coupling of waves and atoms that gives rise
to the new Bose--Einstein statistics. If two particles share the same
phase wave, then they no longer exhibit statistically distinguishable
features. In such a case, there is no meaningful basis to label or track
them as independent entities. From a probability theory perspective,
they must be counted as one wave configuration, not as separate
individuals.

\begin{quote}
If two or more atoms have phase waves that exactly overlap---so that
one may say, consequently, that they are carried by the same
wave---their motions can no longer be regarded as entirely
independent, and these atoms can no longer be treated as distinct
units in probability calculations. The motion of such atoms ``in a
wave'' would thus exhibit a kind of coherence, as a result of
interactions that are impossible to specify precisely, but probably
related to the mechanism that would render unstable the motion of
atoms whose phase wave is not stationary.\footnote{De Broglie (1924, p.~116).}
\end{quote}

Counting particles thus becomes equivalent to counting waves, a shift
that leads de Broglie directly to Bose's derivation of Planck's
radiation law, where the fundamental objects of enumeration are no
longer individual corpuscles but shared wave modes.

From Schrödinger's perspective, de Broglie's thesis had the striking
effect of rendering Einstein's otherwise puzzling gas statistics
conceptually intelligible. What had appeared as an opaque and somewhat
formal manipulation of combinatorics now acquired a clear physical
underpinning through one of the most powerful instruments of classical
mechanics. De Broglie's proposal showed how the statistical anomalies
identified by Einstein could be reinterpreted within a framework that
was not alien to classical theory but rather emerged from its most
refined analytical structures.

At the heart of this reinterpretation lay a profound ontological
reorientation: the shift from a particle-centered to a wave-centered
conception of physical reality. If particles were to be understood as
guided or even constituted by waves, then the new statistics ceased to
be an arbitrary departure from classical reasoning and instead appeared
as the natural consequence of a deeper dynamical picture. What remained
to be done, however, was to transform this ontological insight into a
fully articulated, mathematically coherent mechanics capable of
systematically accounting for quantum phenomena.

\hypertarget{schruxf6dingers-1926-quadrilogy}{%
\subsection{Schrödinger's 1926 Quadrilogy}}

In this section, we examine Schrödinger's celebrated quadrilogy,
concentrating on those elements most directly relevant to our argument.

Schrödinger's first paper on wave mechanics was received by Annalen der
Physik on January 27, 1926, and it is dedicated to the Coulombian
problem of the motion of an electron within the atom.\footnote{Schrödinger (1926a); citations from Schrödinger's papers are
  taken from Schrödinger (1928).} The argument
is both straightforward and obscure, requiring a deep understanding of
de Broglie's thesis to fully grasp. Furthermore, in the previous weeks,
Schrödinger had attempted to derive the relativistic equation without
success. The three-page manuscript containing his calculations reveals
Schrödinger's frustration at not getting anywhere with the relativistic
problem.\footnote{For a discussion of the manuscript, see Duncan and Janssen (2023,
  pp.~392--405).} Thus, Schrödinger begins with the Hamilton-Jacobi equation
for a charge in a field:
\begin{equation}
\frac{1}{2m}|\nabla S|^2-\frac{e^2}{r}=E
\end{equation}

The wave function is then introduced by simply assuming that it is an
exponential function of Hamilton's characteristic function:
\begin{equation}
S=K\ln\psi
\end{equation}

He then substitutes this expression for the wave function into the
Hamilton--Jacobi equation and observes that, whereas in classical
mechanics the equation determines a definite trajectory, in the quantum
case one must instead consider the extremum obtained under small
variations of the wave function. The argument is very brief, with no
detailed justification in the published text. Yet these steps were not
improvised: they rest on a series of prior calculations and exploratory
derivations that Schrödinger had worked out in his private notebooks.
The apparent abruptness of the argument in print thus conceals a more
gradual process of conceptual and mathematical refinement.

The central achievement of Schrödinger's first paper lies in showing
that the imposition of quantum conditions can indeed be
reduced---precisely as the title of the series indicates---to a
classical eigenvalue problem. Starting from the seemingly modest but
physically motivated assumption that the wave function must be finite,
single-valued, and well-behaved throughout configuration space,
Schrödinger demonstrated that the permissible solutions of his equation
form a discrete set of harmonics. These discrete modes of vibration can
be directly correlated with the familiar quantum numbers of the
Bohr--Sommerfeld model. In this way, what had previously appeared as a
somewhat enigmatic feature of the old quantum theory---where quantum
numbers entered through case-to-case prescriptions imposed on classical
orbits---now emerged as a natural mathematical consequence of the
boundary conditions on the wave equation. More fundamentally, this
approach enabled Schrödinger to recast Bohr's frequency condition in an
entirely new light. One striking empirical fact about the old
Bohr-Sommerfeld atomic model was that the frequency of the radiation
emitted during a transition between two stationary states does not
coincide with the frequency associated with either of the individual
orbits. In Bohr's framework, this discrepancy was codified in terms of a
discontinuous ``quantum jump'' between distinct energy levels---a
postulate that, while successful, remained conceptually opaque.

Schrödinger, by contrast, demonstrated that the same phenomenon could be
understood within a continuous wave picture. If atomic states are
described by wave functions, then the transition between two such states
can be represented as the superposition of two undulatory processes.
Their interference gives rise to a new frequency corresponding to the
difference between the original frequencies---a beat frequency. In this
way, what had appeared as an abrupt and discontinuous leap is
reinterpreted as a purely wave-theoretical effect, arising naturally
from the superposition principle and the temporal structure of the wave
function itself.\footnote{Schrödinger (1928, p.~10).}

However, the real conceptual breakthrough happens in the second paper,
which was received on February 23, 1926. In the intervening weeks,
Schrödinger devoted himself with remarkable intensity to the
optical--mechanical analogy. His research notebooks from this period
testify to sustained and often strenuous efforts to clarify the precise
conceptual function of the analogy in the construction of a new
mechanics. Gradually, he came to the conclusion that the analogy, in its
traditional form, was incomplete and required systematic extension.

Hamilton's original formulation had established a correspondence between
classical mechanics and geometrical optics. Yet geometrical optics
itself possesses only a restricted domain of validity: the substitution
of waves by light rays is justified only when diffraction can be
neglected, which is not the case in the domain of quantum phenomena. The
analogy must therefore be carried further. If classical mechanics
corresponds to geometrical optics, then genuine quantum
mechanics---conceived as an extension of classical mechanics into the
microscopic realm---must stand in an analogous relation to wave optics,
see Fig. 1, adapted from Joas and Lehner (2009, 346).

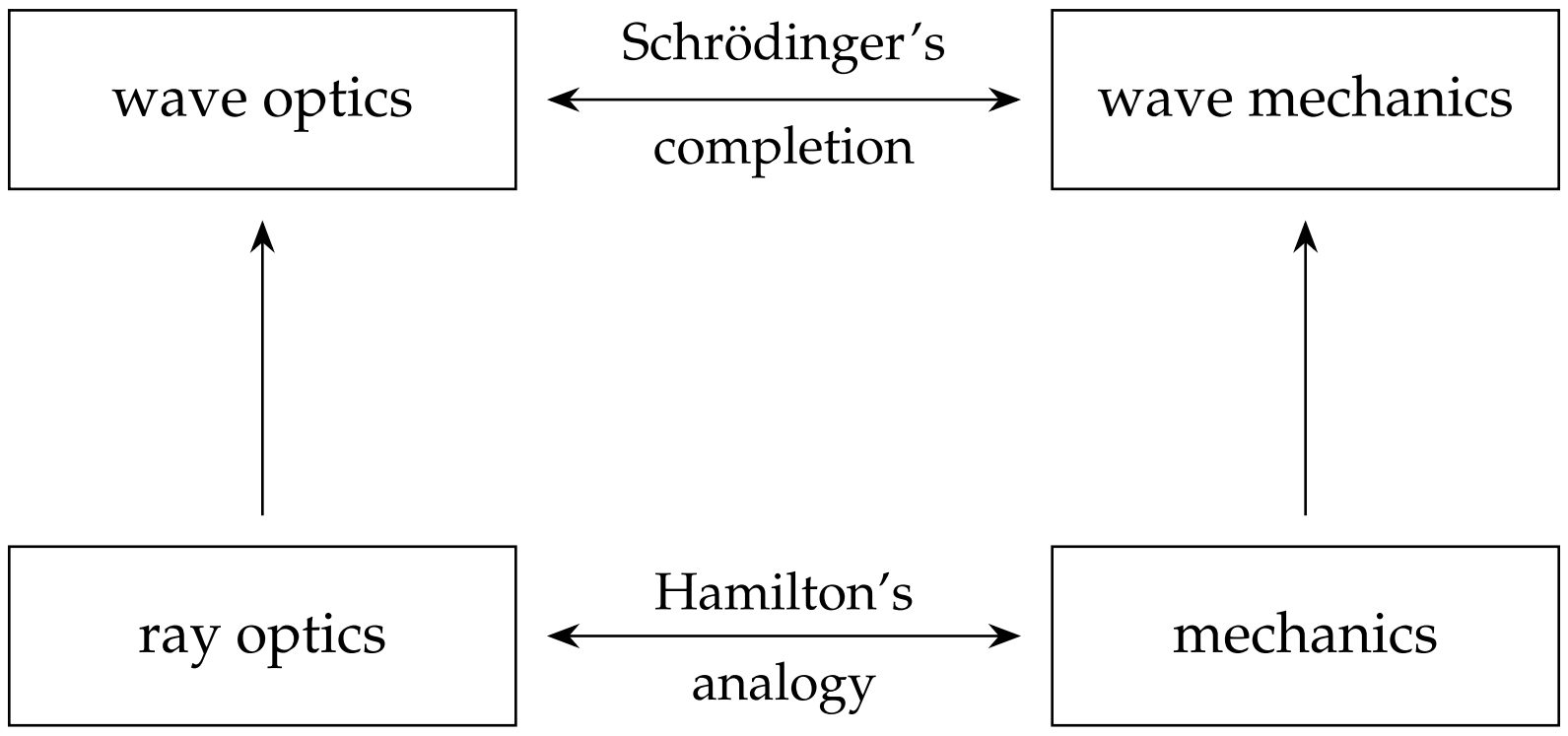
\begin{figure}[t]
\centering
\begin{tikzpicture}[
  box/.style={draw, line width=0.5pt, minimum width=3.4cm, minimum height=1.2cm, text height=1.6ex, text depth=0.4ex, font=\normalsize},
  arr/.style={-{Stealth[length=2.2mm]}, line width=0.5pt},
  lab/.style={font=\small, align=center}]
\node[box] (wo) at (0,3.6) {wave optics};
\node[box] (wm) at (7,3.6) {wave mechanics};
\node[box] (ro) at (0,0) {ray optics};
\node[box] (me) at (7,0) {mechanics};
\draw[arr,{Stealth[length=2.2mm]}-{Stealth[length=2.2mm]}] ([xshift=2mm]wo.east) -- node[lab,above=1pt]{Schrödinger's} node[lab,below=1pt]{completion} ([xshift=-2mm]wm.west);
\draw[arr,{Stealth[length=2.2mm]}-{Stealth[length=2.2mm]}] ([xshift=2mm]ro.east) -- node[lab,above=1pt]{Hamilton's} node[lab,below=1pt]{analogy} ([xshift=-2mm]me.west);
\draw[arr] ([yshift=2mm]ro.north) -- ([yshift=-2mm]wo.south);
\draw[arr] ([yshift=2mm]me.north) -- ([yshift=-2mm]wm.south);
\end{tikzpicture}
\caption{Schrödinger's completion of Hamilton's analogy (picture adapted from Joas and Lehner 2009). When the wavelength is no longer negligible with respect to obstacles and distances between material bodies, geometrical optics must be replaced by wave optics. Analogously, particle mechanics must be replaced by wave mechanics when the wavelength is comparable to atomic distances (see Schrödinger 1928, p.~25).}
\label{fig:analogy}
\end{figure}

This crucial insight is formulated in the second paper:

\begin{quote}
We know to-day, in fact, that our classical mechanics fails for very
small dimensions of the path and for very great curvatures. Perhaps
this failure is in strict analogy with the failure of geometrical
optics, i.e.~``the optics of infinitely small wave lengths'', that
becomes evident as soon as the obstacles or apertures are no longer
great compared with the real, finite, wave length. Perhaps our
classical mechanics is the complete analogy of geometrical optics and
as such is wrong and not in agreement with reality; it fails whenever
the radii of curvature and dimensions of the path are no longer great
compared with a certain wave length, to which, in q-space, a real
meaning is attached. Then it becomes a question of searching for an
undulatory mechanics, and the most obvious way is the working out of
the Hamiltonian analogy on the lines of undulatory optics.\footnote{Schrödinger (1928, p.~18).}
\end{quote}

The second paper is structured in two distinct parts. In the first,
Schrödinger articulates his reinterpretation of the optical--mechanical
analogy and clarifies its foundational significance for wave mechanics.
In classical mechanics, the solutions of the Hamilton--Jacobi equation
define surfaces of constant action, which in turn determine the possible
motions of the system. In this new approach, Hamilton's characteristic
function is seen as a phase function, and the surfaces of constant
action are understood as wave fronts associated with a mechanical point.

On this basis, Schrödinger reverses the traditional hierarchy between
particles and waves. The wave is treated as ontologically primary, while
what are ordinarily described as particles are reconceived as localized
peaks arising from the superposition of waves propagating through space.
In this way, the optical--mechanical analogy is no longer a merely
formal correspondence but becomes the conceptual bridge toward a
fundamentally wave-based account of physical reality:

\begin{quote}
What I now categorically conjecture is the following: The true
mechanical process is realised or represented in a fitting way by the
\emph{wave processes} in q-space, and not by the motion of \emph{image points}
in this space. The study of the motion of image points, which is the
object of classical mechanics, is only an approximate treatment, and
has, as such, just as much justification as geometrical or ``ray''
optics has, compared with the true optical process. A macroscopic
mechanical process will be portrayed as a wave signal of the kind
described above, which can approximately enough be regarded as
confined to a point compared with the geometrical structure of the
path. We have seen that the same laws of motion hold exactly for such
a signal or group of waves as are advanced by classical mechanics for
the motion of the image point. This manner of treatment, however,
loses all meaning where the structure of the path is no longer very
large compared with the wave length or indeed is comparable with it.
Then we must treat the matter strictly on the wave theory, i.e., we
\emph{must} proceed from the \emph{wave equation} and not from the fundamental
equations of mechanics, in order to form a picture of the manifold of
the possible processes. These latter equations are just as useless for
the elucidation of the micro-structure of mechanical processes as
geometrical optics is for explaining the \emph{phenomena of
diffraction}.\footnote{Schrödinger (1928, p.~25).}
\end{quote}

At this stage in his intellectual development, Schrödinger attributed a
strong ontological significance to the emerging formalism. Wave
mechanics was, in his view, a genuine extension of classical mechanics
into the quantum domain. Rather than abandoning the conceptual
foundations of the classical worldview, he sought to preserve and
generalize them. In this sense, wave mechanics appeared to him as a
natural continuation of the classical picture of reality---one capable
of accommodating quantum phenomena without relinquishing the ideal of a
coherent, continuous, and intelligible physical world. This view is even
more apparent when he discusses the status of de Broglie's phase wave,
to which he attributes properties very familiar to modern ears:

\begin{quote}
In \emph{this} sense do I interpret the ``phase waves'' which, according to
de Broglie, accompany the path of the electron; in the sense,
therefore, that no special meaning is to be attached to the electronic
path itself (at any rate, in the interior of the atom), and still less
to the position of the electron on its path. And in this sense I
explain the conviction, increasingly evident to-day, \emph{firstly}, that
real meaning has to be denied to the phase of electronic motions in
the atom; \emph{secondly}, that we can never assert that the electron at a
definite instant is to be found on any definite one of the quantum
paths, specialised by the quantum conditions; and \emph{thirdly}, that the
true laws of quantum mechanics do not consist of definite rules for
the single path, but that in these laws the elements of the whole
manifold of paths of a system are bound together by equations, so that
apparently a certain reciprocal action exists between the different
paths.\footnote{Schrödinger (1928, p.~26).}
\end{quote}

In the second part of the article, Schrödinger applies his approach to
some long-standing problems of old quantum theory, such as Planck's
oscillator, the rotator, and the dumbbell model.

The third paper is devoted primarily to the development and application
of perturbation theory, which Schrödinger employs to calculate atomic
spectra in the presence of external electric or magnetic fields. In this
context, wave mechanics proves its technical power by reproducing and
extending results previously obtained within the old quantum theory,
while offering a more systematic and unified method of approximation.
For the purposes of our argument, however, the fourth paper---received
on June 23, 1926---is of greater significance. In this final installment
of the series, Schrödinger addresses issues that go beyond computational
technique and bear directly on the conceptual interpretation of the new
theory. Two elements, in particular, deserve closer attention. First, to
properly treat the phenomenon of dispersion and to retrieve Kramers'
theory, Schrödinger is forced, for the first time, to come to terms with
the fact that the wave function is complex-valued, i.e.~has a real and
an imaginary part. In a letter to Wilhelm Wien on June 18, 1926,
Schrödinger refers to the wave function as a ``scalar field.'' A few days
later, in the fourth paper, he presents this interpretation as a
``heuristic hypothesis,'' which entails that the product between the wave
function and its complex conjugate, multiplied by the charge of the
particle (he was considering the wave function of an electron),
``represents the electrical density as a function of the space
co-ordinates and the time'' (Schrödinger 1928, 108). Thus, although
Schrödinger accepts the complex aspects of his theory, he remains
determined to maintain contact with the classical picture in space and
time. This interpretation will be examined in detail in the next
section.

\hypertarget{schruxf6dingers-view-of-the-wave-function}{%
\section{Schrödinger's View of the Wave Function}}

The wave mechanics formulated by Schrödinger in his four communications,
in contrast with the matrix mechanics developed by Heisenberg, Born and
Jordan, seemed to bring some insight into the description of the
microscopic processes. Quantum systems were now described by the wave
function, a complex scalar field that looked like a physical wave of
classical mechanics. In analogy with wave optics, Schrödinger
interpreted the wave function as a material wave composing the
microscopic particles.\footnote{More precisely, it was the absolute square of the wave function
  to have a material interpretation, as we will analyze later.} In this view, a microscopic particle
evolving in time would correspond to a well-localized wave packet moving
with group velocity. Schrödinger intended, indeed, wave mechanics as a
generalization of particle mechanics, in analogy with the relation
between wave optics and geometrical optics (Fig. 1). Following this
intuition, it was natural to interpret quantum theory as (i) a theory of
physical waves propagating in 3D space and (ii) a deeper theory with
respect to particle mechanics having the latter as a limiting case in
the appropriate (macroscopic) conditions.

\hypertarget{lorentzs-objections}{%
\subsection{Lorentz's objections}}

The first objections to Schrödinger's realist interpretation of the wave
function were raised by Lorentz in a lively letter exchange.\footnote{The letter exchange between Lorentz and Schrödinger is published
  in the volume edited by Przibram (1967).} These
can be grouped in two categories:

\begin{enumerate}
\def\labelenumi{\arabic{enumi}.}
\item
  Contrary to classical waves, the wave function is defined in the
  system's configuration space. For an
  \(n\)-particle system, this means the
  wave function is defined and evolves in a space of
  \(3n\) dimensions. This entity cannot be
  thought of as a physical entity.
\item
  A wave packet is typically unstable and spreads out in a very short
  time. A particle instead is a stable and well-localized entity, thus
  it cannot be thought of as materially composed of the wave function.
\end{enumerate}

We note that these two objections, raised by Lorentz in 1926, are still
the main objections against those approaches which interpret the wave
function in terms of a physical object.\footnote{Modern solutions to the configuration space problem fall into two
  groups: one option is to accept that the wave function describes a
  new object in configuration space, and assign to configuration space
  the role of fundamental physical space. This option is known as wave
  function realism (Albert 1996; Ney 2021; see also Calosi, this
  volume, for a taxonomy of the varieties of wave function realism). A
  second option is to follow the route initiated by Schrödinger,
  defining a novel object in three-dimensional space derived from the
  wave function. This is the case with the mass-density interpretation
  formulated by Ghirardi et al.~(1995) and the polywave/multi-field
  interpretation originally formulated by Forrest (1988), discussed by
  Belot (2012) in the context of Bohm's theory and defended by Hubert
  \& Romano (2018) and Romano (2021) (see also Rusi El Hassani and
  López, this volume, for a discussion of the multi-field view in
  connection with separability). We will discuss the mass-density and
  the multi-field approaches in sect. 5.}

Both objections are raised by Lorentz in the first letter to Schrödinger
(27 May 1926). The problem of configuration space appears at the end of
the first page:

\begin{quote}
If I had to choose now between your wave mechanics and the matrix
mechanics, I would give the preference to the former, because of its
greater intuitive clarity, \emph{so long as one only has to deal with the
three coordinates}
\(x,y,z\). If, however, there are more degrees of
freedom, then I cannot interpret the waves and vibrations physically,
and I must therefore decide in favor of matrix mechanics.\footnote{Przibram (1967), letter 19: Lorentz to Schrödinger (emphasis
  added).}
\end{quote}

According to Lorentz, the fact that the wave function is defined in
configuration space is a sign that it cannot be interpreted physically.
Thinking of a physical entity in configuration space was not a
reasonable option at the time either for Lorentz or for Schrödinger. If
the wave function was a physical object, then it had to be defined
(somehow) in 3D space (on the debate about the status of configuration
space in early quantum mechanics, see also Allori, this volume).

A few pages later, Lorentz expresses his second objection, concerning
the identification of a material particle with a wave packet proposed by
Schrödinger:\footnote{Ibidem.}

\begin{quote}
Your conjecture that the transformation which our dynamics will have
to undergo will be similar to the transition from ray optics to wave
optics sounds very tempting, but I have some doubts about it.

If I have understood you correctly, then a ``particle'', an electron for
example, would be comparable to a wave packet which moves with the
group velocity.

But a wave packet can never stay together and remain confined to a
small volume in the long run. The slightest dispersion in the medium
will pull it apart in the direction of propagation, and even without
that dispersion it will always spread more and more in the transverse
direction. Because of this unavoidable blurring a wave packet does not
seem to me to be very suitable for representing things to which we
want to ascribe a rather permanent individual existence.
\end{quote}

The first paragraph of this passage explains why Schrödinger aimed for
such identification between a microscopic particle and a wave packet. As
mentioned before, Schrödinger regarded quantum mechanics as a
generalization of particle mechanics in the microscopic regime. The
transition from particle mechanics to wave mechanics was, according to
Schrödinger, analogous to the transition from ray optics to wave optics.
As ray optics is a limiting case of wave optics when diffraction effects
can be neglected, particle mechanics was a limiting case of wave
mechanics when the ``wave effects'' could be neglected. That is: when the
wave function is well-localized and remains well-localized in time, this
will appear as a discrete and corpuscular object in the macroscopic
regime, i.e.~it will appear as a particle.

The objection raised by Lorentz was a serious one: even if we can
identify a ``particle'' with a wave packet at the initial time, the wave
packet will tend to spread out and therefore will not maintain a compact
structure for a long time (the precise time depending on the state and
the external potential). This makes the identification with the particle
difficult to maintain.\footnote{There are exceptions: stationary states (atomic orbitals) and
  coherent states (harmonic oscillator) remain well-localized.
  However, this is not valid in general and typically the wave
  function tends to spread out in time.}

In a second passage Lorentz refines his previous objection. This time he
is concerned that Schrödinger's interpretation could not explain the
experimental results coming from the photoelectric effect: when
electrons are extracted from metals, they concretely appear as discrete
particles and not as distributed waves. If electrons were to be regarded
as physical waves constituting atomic orbitals, how was it possible for
them to transform into particles when extracted from metals?\footnote{Ibidem.}

\begin{quote}
The disadvantage, and it is indeed a serious one, is this: whatever we
assume about the electron in the hydrogen atom we must also assume for
all electrons in all atoms: we must replace them all by systems of
waves. But then \emph{how am I to understand the phenomena of
photoelectricity and the emission of electrons from heated metals?}
The particles appear here quite clearly and without alteration; once
dissolved, how could they condense again?
\end{quote}

This second problem raised by Lorentz can be thought, in modern terms,
as the problem to describe emergent discrete particles starting from
distributed waves. In the next subsection we will discuss a possible
solution to this problem, which involves decoherence theory and
non-standard interpretations of quantum mechanics. Lorentz with this
objection was pointing at one of the major foundational problems of the
theory, i.e.~the classical limit of quantum mechanics. Independently of
the specific interpretation of the wave function, in fact, the problem
of describing particles or (more generally) classical objects from
quantum theory in the macroscopic regime is still an open problem in
many respects.\footnote{Whereas decoherence theory explains much of this transition, it
  does not explain all. Moreover, there is a large consensus nowadays
  that decoherence should be implemented in realist interpretations of
  quantum mechanics to provide a classical ontology. The debate on
  this problem, in our opinion, is far from being settled (see Crull
  2013 and Schlosshauer 2019 for reviews of decoherence theory, Romano
  2022 for the limits of decoherence in standard quantum mechanics,
  and Bacciagaluppi 2025 for an overview of decoherence in
  non-standard theories).}

\hypertarget{solving-the-wave-packet-problem}{%
\subsection{Solving the ``wave packet'' problem}}

Concerning the wave packet problem, Schrödinger was interested most of
all in describing atomic orbitals, i.e.~stationary states. However,
Lorentz was correct in pointing out that this strategy could not be
extended \emph{simpliciter} to general cases. An interesting question is to
ask how this problem is solved today. Overall, there are two different
strategies to attack this problem nowadays: either we use decoherence;
or non-standard theories (plus decoherence).\footnote{By ``non-standard theories'' we refer here to the GRW theory,
  Bohm's theory and Everett theory.} Authors who are
inclined to see the problem solved by standard decoherence\footnote{By ``standard decoherence'' we refer to decoherence theory in the
  standard quantum mechanics.} stress
the fact that the interaction with the external environment leads to a
progressive localization in the position basis (for macroscopic systems)
and this results in the selection of Gaussian pointer states (Zurek et
al.~1993). In turn, these selected states can be interpreted as
particles in the macroscopic regime (e.g.~Zeh (1993)). The localization
process is mathematically represented by the progressive diagonalization
of the subsystem's reduced density matrix,\footnote{In decoherence, the system of interest is (after the interaction
  with the external environment) a subsystem of the system-environment
  entangled state. The subsystem cannot be represented by a wave
  function (being part of a larger entangled state), so it is
  represented by a reduced density matrix. This is a further problem
  of standard decoherence: there is no wave function for pointer
  states, and no wave function for emergent classical objects.} which makes quantum
interference (represented by the non-diagonal terms of the matrix)
between different subsystem's relative states empirically not
detectable.\footnote{Zurek (1991). Schlosshauer (2007, 2019).} However, the localization process of decoherence does
not reduce the subsystem to a single localized state, but to many
localized states,\footnote{The term ``localization'' is a misnomer associated with the
  decoherence effect. ``Localization'' is intuitively associated with
  the idea of a unique localized position, while the localization
  process of decoherence leads the (sub)system to a (reduced) state
  represented by different possible localized positions. No one of
  these positions is truly selected until a measurement is performed.
  A term like ``multi-localization'' would be more precise.} all represented by the diagonal terms of the
reduced density matrix.\footnote{See e.g.~Romano (2022) for an analysis of the localization
  process and the interpretation of decoherence.} Concretely, if a system is initially
described by a superposition of two Gaussians (e.g.~double-slit
experiment), the interaction with the environment will reduce the system
to two (non-overlapping, non-interfering) Gaussians, not one localized
Gaussian. Performing a double slit experiment with decoherence, we will
see the formation of a classical probability distribution of the
electrons on the photographic plate, but we cannot associate individual
trajectories to the electrons arriving at the plate (which should be
possible instead in case of a real reduction of the initial
superposition to a unique localized state).

The other strategy to solve the wave packet problem is to opt for a
realist non-standard theory, that is, one among the spontaneous collapse
models (Bassi \& Ghirardi 2003), Bohm's theory (Bohm 1952) or the Many
Worlds Interpretation (Wallace 2012). In particular, both in Bohm's
theory and in the spontaneous collapse models the wave function will
typically be well-localized in the macroscopic limit or in measurement
interactions: in the former theory this is due to the effective
factorization process,\footnote{See e.g.~Bohm et al.~(1987), Romano (2023).} in the latter to the amplification mechanism
of the spontaneous collapse rate for systems having many degrees of
freedom.\footnote{Bassi \& Ghirardi (2003). This aspect will be analyzed in sect. 5
  presenting the mass-density interpretation.}

\hypertarget{schruxf6dingers-electric-charge-density-solving-the-configuration-space-problem}{%
\subsection{Schrödinger's electric charge density: solving the configuration space problem}}

Schrödinger was aware that the wave function, being defined in
configuration space, could not be directly interpreted as a physical
field. Also, he tended to regard the wave function as a formal tool,
since it is complex-valued. To overcome this problem, Schrödinger tried
to formulate a new dynamical equation for a real-valued wave function.
After various attempts, he ended up with a real-valued, fourth order,
differential equation formally similar to the equation of a classical
vibrating plate:
\begin{equation}
\hbar^2\frac{\partial^2\psi}{\partial t^2}=-H^2\psi
\end{equation}

This equation, however, was more difficult to use than the Schrödinger
equation for the complex-valued wave function, and presented technical
difficulties that did not arise in the original equation.\footnote{A historical and philosophical analysis of Schrödinger's fourth
  order, real-valued, differential equation can be found in Callender
  (2023) and references therein.} For these
reasons it was eventually abandoned.

Even if the wave function was complex-valued, Schrödinger realized that
the absolute square of the wave function
\(\psi^*\psi=|\psi|^2\) was a real-valued function, thus
representing a real scalar field. He readily adopted this quantity to
specify the physical meaning of the wave function. In particular,
Schrödinger interpreted the absolute square of the wave function
multiplied by the charge of an electron, i.e.~the quantity:
\begin{equation}
-e|\psi|^2 \tag{\ref{eq:charge}}
\end{equation}
as the \emph{electron charge density}, i.e.~a continuous real-valued scalar
field that specified the intensity of the charge at each point in 3D
space. It was Schrödinger, not Born, who first assigned the physical
significance of the wave function to its absolute square.\footnote{The shift from a realist to an instrumental understanding of
  quantum mechanics can be largely summarized as the transition from
  the absolute square as a material field as intended by Schrödinger
  to the absolute square as a probability density as intended by Born.
  When the latter interpretation is adopted, the instrumental
  character of the formalism just follows since the wave function
  ceases to describe anything real before the measurement.} Moreover,
and most importantly, Schrödinger found a way to define the electron
charge density in 3D space even for
\(n\)-particle systems. The main idea was to
think of the total charge density as the sum of the individual charge
densities of each electron, with the latter mathematically defined in 3D
space. Schrödinger expresses this idea in a letter to Lorentz (6 June
1926):\footnote{Przibram (1967), letter 20: Schrödinger to Lorentz, pp.~61--62.}

\begin{quote}
You mention the difficulty of projecting the waves in q-space, when
there are more than three coordinates, into ordinary three-dimensional
space and of interpreting them physically there. I have been very
sensitive to this difficulty for a long time but believe that I have
now overcome it. I believe {[}...{]} that the physical meaning belongs
not to the quantity itself but rather to a \emph{quadratic} function of it.
\emph{There} I chose the real part of
\(\psi\bar\psi\), where $\psi$ is taken to be complex in
the obvious way {[}...{]} and the bar denotes the complex conjugate.
\emph{Now} I want to choose more simply
\(\psi\bar\psi\), that is, the square of the absolute
value of the quantity $\psi$. If we now have to deal with N particles, then
\(\psi\bar\psi\) (just as $\psi$ itself) is a function of 3N
variables or, as I want to say, of N three dimensional spaces, R1, R2,
\ldots, RN. Now first let R1 be identified with the real space and
integrate \(\psi\bar\psi\) over R2, \ldots, RN; second, identify R2
with the real space and integrate over R1, R3, \ldots, RN; and so on. The
N individual results are to be added after they have been multiplied
by certain constants which characterize the particles, (their charges,
according to the former theory). I consider the result to be the
electric charge density in real space.
\end{quote}

A few weeks later this idea appeared in print in Schrödinger's fourth
communication on wave mechanics:\footnote{Schrödinger (1928), sect. 7: ``On the physical significance of the
  field scalar'', pp.~120--123.}

\begin{quote}
The heuristic hypothesis of the electro-dynamical meaning of the field
scalar \(\psi\), previously employed in the
\emph{one}-electron problem, was extended off-hand to an arbitrary system
of charged particles in § 2, and there a more exhaustive description
of the procedure was promised. We had calculated the density of
electricity at an arbitrary point in space as follows. We selected
\emph{one} particle, kept the trio of co-ordinates that describes \emph{its}
position in ordinary mechanics fixed; integrated
\(\psi\bar\psi\) over all the rest of the co-ordinates
of the system and multiplied the result by a certain constant, the
``charge'' of the selected particle; we did a similar thing for each
particle (trio of co-ordinates), in each case giving the selected
particle the same position, namely, the position of that point of
\emph{space} at which we desired to know the electric density. The latter
is equal to the algebraic sum of the partial results.
\end{quote}

The description of the procedure is clear as it was in the letter to
Lorentz: computing the ``density of electricity'' for each particle and
summing all the partial contributions (the contributions to the total
charge given by each particle) to calculate the total charge density.
What was relevant is that integrating out the position coordinates of
the other particles when considering a specific particle allowed a
precise definition of the charge density in 3D space. This was a way to
connect an abstract complex-valued entity defined in configuration
space, the wave function, to a physical real-valued entity defined in
ordinary 3D space, the electron charge density. The latter could then be
interpreted as a real continuous field composing the microscopic
objects.

Concerning the wave function \emph{per se}, it seems to have no meaning other
than that of a formal tool to compute the real ``stuff'' in ordinary
space, i.e.~the continuous field of the electric charge density. In many
passages Schrödinger stresses this point: he does not see the wave
function as a real object, but as a formal tool to define the electron
charge density:\footnote{Schrödinger (1928, p.~120)}

\begin{quote}
This rule is now equivalent to the following conception, which allows
the true meaning of
\(\psi\) to stand out more clearly.
\(\psi\bar\psi\) is a kind of \emph{weight -function} in the
system's configuration space. The \emph{wave-mechanical} configuration of
the system is a \emph{superposition} of many, strictly speaking of all,
point-mechanical configurations kinematically possible. Thus, each
point-mechanical configuration contributes to the true wave-mechanical
configuration with a certain \emph{weight}, which is given precisely by
\(\psi\bar\psi\). If we like paradoxes, we may say that
the system exists, as it were, simultaneously in all the positions
kinematically imaginable, but not ``equally strongly'' in all.
\end{quote}

There are two important points in this passage:

\begin{enumerate}
\def\labelenumi{\arabic{enumi}.}
\item
  The absolute square of the wave function (without the electron
  charge) is interpreted as a \emph{weight-function}, i.e.~a function that
  ``weights'' the density of electricity for each particle and
  eventually gives the total density for the global (many particle)
  system. This is in stark contrast with the interpretation of the
  absolute square as \emph{probability density} given by Born. Whereas the
  notion of a \emph{weight-function} indicates something real to which the
  weight is assigned, the notion of \emph{probability density} is connected
  to the probabilities of the measurement outcomes, leading the way to
  an instrumental interpretation of quantum mechanics.
\item
  Schrödinger describes the total ``wave-mechanical configuration'' in
  space for an
  \(n\)-particle system as a superposition
  of different ``point-mechanical configurations''. Saying that the
  former is a superposition (in physical space) of the latter
  establishes a precise idea of the relation between configuration
  space and physical space. That is: configuration space is a formal
  tool to describe real superpositions in physical space.
\end{enumerate}

The interpretation of the wave function proposed by Schrödinger, i.e.~to
regard $-e|\psi|^2$ as the electric charge density in 3D
space, survived (directly or indirectly) into the modern realist
approaches to the wave function. We will discuss in section 5 two of
these interpretations: the \emph{mass-density} approach proposed by Ghirardi
et al.~(1995) and the polywave/multi-field approach proposed by Forrest
(1988) and later defended by Hubert \& Romano (2018) and Romano (2021).
Before turning to the realist approaches, however, in the next section
we will take a detour into quantum chemistry, where Schrödinger's
interpretation still proves useful.

\hypertarget{sebens-charge-density-and-quantum-chemistry}{%
\section{Sebens: charge density and quantum chemistry}}

In a recent paper, Sebens (2021) argues that the interpretation of the
electron as a continuous charge density (rather than a point-like
charged particle, as it would be suggested by Born's interpretation)
finds strong support in quantum chemistry (on this topic, see also
Lombardi and Fortin, this volume). The argument is that two of the most
important techniques used in quantum chemistry to evaluate the shape of
the molecules, namely the Hartree-Fock method and the Density Functional
Theory, implicitly interpret the atomic orbitals as a continuous
electron charge density. More than that, in the calculations that are
employed in these techniques the atomic orbitals are already described
as charge densities.

In particular, the Hartree--Fock method approximates the ground-state
potential energy of an atom or a molecule as the sum of five
contributions (Sebens 2021, p.~14):

\begin{quote}
1. Nucleus-Nucleus Repulsion

2. Minus Electron-Nucleus Attraction

3. Plus Electron Repulsion

4. Minus Electron Self-Repulsion

5. Minus Exchange Energy
\end{quote}

The first three contributions can be interpreted as the classical
electrostatic potential energy associated with the continuous electron
charge density interacting with the (point-like) nuclei, while the other
two contributions describe quantum corrections. An objection can be made
to the extent that the electron charge density \emph{per se} is not able to
explain all contributions to the total energy. This is discussed by
Sebens in the following passage:\footnote{Sebens (2021, pp.~14--15).}

\begin{quote}
I think this gives us strong reason to take Schrödinger's proposal
seriously. One might respond that Schrödinger's charge distribution is
inadequate because it fails to give a classical explanation of all
five contributions to the potential energy. There are ways the details
might be filled in so that it succeeds in doing so {[}...{]}, but I
don't see why it must. It is not surprising that the expression for
energy in quantum physics is different from classical
electromagnetism. The fact that the classical energy of Schrödinger's
charge distribution appears as a contribution to the total energy
makes it appealing to say that we have familiar stuff (spread-out
matter with charge) obeying new rules (quantum physics).
\end{quote}

In other words, if the electrons were point-like charged particles with
definite but unknown positions, the classical electrostatic
interpretation of these energy terms would be difficult to make.
Instead, in that case (point-like electrons) one would have to
reinterpret these terms as expectation values over hypothetical
configurations, which fails to explain why the classical electrostatic
structure appears so directly in the energy decomposition.

The Density Functional Theory (DFT), like the Hartree-Fock method, is a
method used in quantum chemistry to approximate various properties of
atoms and molecules. In the DFT the total energy of the atom or the
molecule is expressed as a functional of the electron charge density
(differently interpreted in different textbooks), the latter being a
function of position. Leaving aside the technical details of the
method,\footnote{For the interested reader, see Sebens (2021, sect. 3).} also in this case some of the terms appearing in the
equations can be explained (or naturally explained) only interpreting
the electrons as a continuous distribution of charge density and not as
charged particles. The fact that both these techniques employ the notion
of electric charge density or, more precisely, the fact that some of the
contributions in the mathematical equations used to calculate the form
of the atoms/molecules are naturally interpreted as the result of the
electromagnetic potential energy associated with a continuous charge
density gives support to the original interpretation proposed by
Schrödinger. Understanding why molecular structure emerges as it does
requires interpreting $|\psi|^2$ not merely as a probability density but as
encoding an actual distribution of electric charge. Without this
interpretation, the classical contributions to the atoms and molecules
potential energy in Hartree--Fock and DFT appear formally coincidental
rather than physically explanatory. Sebens convincingly concludes
that:\footnote{Sebens (2021, p.~24).}

\begin{quote}
In approximating the ground state energy and {[}the{]} electron
structure of atoms and molecules, we have seen that both the
Hartree-Fock method and density functional theory contain clues that
electron charge really is spread-out in the way Schrödinger suggested.
{[}...{]} Seeing how chemists approximate the underlying quantum
physics of atoms and molecules helps to explain why chemists often
apply Schrödinger's role for amplitude-squared even though Born's role
is generally viewed as having superseded it.
\end{quote}

\hypertarget{from-the-wave-function-to-matter-in-3d-space}{%
\section{From the wave function to matter in 3D space}}

The interpretation of the wave function as a tool for defining a
material field in 3D space has been taken up lately, directly or
indirectly, in other realist interpretations of the wave function.
Directly with the notion of the mass-density in the spontaneous collapse
models, which is analogous to Schrödinger's electron charge density but
implemented with a different property, the mass of the particles, and in
a different framework, which solves the ``wave packet'' problem.
Indirectly with the notion of ``polywave'' proposed by Forrest (1988) in
standard quantum mechanics and the multi-field view defended by Hubert \&
Romano (2018) in Bohm's theory. The structure of the
polywave/multi-field proposal is different in some respect from
Schrödinger's original idea: in the former the wave function is not a
formal tool to define a physical entity in 3D space, but it actually
represents a physical entity in 3D space. Still, we believe that also
this proposal includes an important part of Schrödinger's insight, i.e.
the idea that configuration space is just a mathematical space and,
whatever exists in the ontology according to quantum theory, this must
be defined as a physical entity in 3D space. We will analyze these two
realist interpretations in the next two subsections.

\hypertarget{the-mass-density-interpretation}{%
\subsection{The mass-density interpretation}}

With the term ``spontaneous collapse models'' of quantum mechanics we
include a family of theories in which the Schrödinger evolution is
modified by non-linear and stochastic terms which trigger spontaneous
collapses of the wave function at random times and random
locations.\footnote{For a complete presentation of dynamical reduction models see
  e.g.~Bassi \& Ghirardi (2003).} The frequency of the spontaneous collapse depends on the
model but in all models it is very unlikely for microscopic systems (so
as to maintain quantum superpositions) and very likely and effective for
macroscopic systems (so as to make superpositions disappear in the
classical regime). The spontaneous collapse can be discrete, as in the
Ghirardi-Rimini-Weber (GRW) theory (Ghirardi et al.~1986), or continuous
as in the Continuous Spontaneous Localization (CSL) model (Pearle 1989;
Ghirardi, Pearle \& Rimini 1990). The main idea of the spontaneous
collapse models is to have a unified dynamics for the microscopic and
the macroscopic regime (i.e.~the same dynamics describing photons or
electrons in the microscopic regime gives rise to classical behavior
when the system considered is composed of many degrees of freedom).
Moreover, as in Bohm's and Everett's theories, also the spontaneous
collapse models do not have a special postulate for describing
measurement interactions, and the measurement problem is solved just by
the dynamics, which eliminates (through the spontaneous collapse) linear
superpositions between different macroscopic configurations.

The first spontaneous collapse model was introduced by Ghirardi and
collaborators in 1986 and is known as the GRW theory. In this theory, a
microscopic particle evolves linearly most of the time and undergoes a
spontaneous collapse with frequency
\(\lambda=10^{-16}\,\mathrm{s}^{-1}\). The spontaneous collapse, described by
a ``localization operator'', is mathematically represented by a
multiplication with a Gaussian that localizes the system and eliminates
superpositions. The evolution of a microscopic system in the GRW theory
is thus represented by a linear dynamics interrupted by discrete ``hits''
or ``localizations'' at random times. For macroscopic systems instead the
situation changes drastically, since the frequency of the ``hits''
increases proportionally with the number of particles considered (this
is known as the ``amplification mechanism''). Concretely, for an
\(n\)-particle system the collapse frequency
is \(n\lambda\), so that a macroscopic system composed of a number of
particles proportional to
\(n=10^{23}\) is always well-localized. If, on the one
hand, the GRW theory provides a precise dynamics for micro and macro
systems and solves the measurement problem, on the other hand, it does
not provide a clear ontology for the microscopic world. The GRW theory,
as standard quantum mechanics, is fundamentally a theory about the wave
function, a complex-valued scalar field defined in configuration
space.\footnote{To a certain extent also standard quantum mechanics is a theory
  about the wave function, but theories like GRW were formulated
  exactly to solve the philosophical problems of the standard theory,
  such as the measurement problem and the ontology problem. Not all
  physicists embracing the spontaneous collapse models think however
  that an explicit ontology should be combined with the formalism of
  GRW, as the theory already presents a precise description of the
  dynamics and all possible interactions between systems. It is mostly
  thanks to the philosophical sensibility of Ghirardi, and his
  continuous exchange with J. S. Bell, that an explicit ontology for
  the spontaneous collapse models was eventually proposed.} How is this field supposed to describe microscopic particles
and, eventually, macroscopic objects such as tables and chairs in
three-dimensional space?

To solve this problem, Ghirardi and collaborators (Ghirardi et al.~1995)
introduced the notion of mass-density, i.e.~a model in which matter is
described by a continuous mass-density field in ordinary 3D space. In
particular, the mass-density interpretation is directly implemented in
the Continuous Spontaneous Localization model in which, differently from
the GRW theory, the wave function evolves according to a continuous and
linear stochastic evolution (eq. (2.6) in Ghirardi et al.~(1995)). The
value of the mass-density at point
\(x\) and time
\(t\) is defined as
\begin{equation}
m(x,t)=\langle\psi_t|\hat M(x)|\psi_t\rangle
\end{equation}
where \(\psi_t\) is the stochastically evolving wave
function and \(\hat M(x)\) the mass density operator at position
\(x\). Written in explicit form, the mass
density operator for an \(n\)-particle system is:
\begin{equation}
\hat M(x)=\sum_{i=1}^{N}m_i\,\delta(\hat x_i-x)
\end{equation}
where \(m_i\) and
\(\hat x_i\) are respectively the mass and the
position operator of the particle
\(i\) and
\(\delta\) is the Dirac delta function. So the mass-density field
becomes:
\begin{equation}
m(x,t)=\sum_{i=1}^{N}m_i\int|\psi_t(x_1,\dots,x_N)|^2\,\delta(x_i-x)\,dx_1\dots dx_N
\end{equation}
which is analogous to the electron charge density defined by Schrödinger
(for a many particle system of \emph{n} electrons with wave function
\(\psi(x_1,\dots,x_N)\)):\footnote{For the derivation and definition of the electron charge density
  in modern mathematical terms see e.g.~Sebens (2021, pp.~6--7) or
  Norsen (2017, p.~130, in particular eqs. (5.25)--(5.26)).}
\begin{equation}
\rho_c(x,t)=\sum_{i=1}^{N}q_i\int|\psi(x_1,\dots,x_N)|^2\,\delta(x_i-x)\,dx_1\dots dx_N
\end{equation}
where \(\rho_c\) is the total electron charge density, and
\(q_i=-e\) is the charge of a single electron.
That is: the notion of mass density is just the original definition by
Schrödinger but this time implemented with the mass of the particles
instead of the electric charge. Ghirardi and collaborators explain in
the paper that they prefer to connect the ontology of the theory to the
mass of the particles for some important reasons: first of all, the
(smeared) mass density operator defines the dynamical equation of the
CSL model, in this way the mass density becomes an integral part of the
formalism of the theory and not something added to the wave function.
Secondly, the authors wanted to connect the stochastic collapses to the
property of the mass in view of the fact that a deeper origin for the
spontaneous collapse might be traced back to gravity. Furthermore, the
mass-density field is generated also by electrically neutral particles.
Therefore, the mass density can be thought of as a generalization and
refinement of the original electric charge density:

\begin{itemize}
\item
  \emph{Generalization}: since it applies to charged and neutral particles
\item
  \emph{Refinement}: since the mass-density is implemented in the dynamics
  of the CSL model and the continuous collapse processes keep the mass
  density localized for macroscopic systems, solving the major problem
  of the original proposal
\end{itemize}

According to Ghirardi and collaborators, the mass-density is the
ontology of the theory: it is what atoms and molecules, tables and
chairs are made of, what exists in the micro and macro world according
to quantum theory. This idea, i.e.~that the wave function is a formal
tool which guides the evolution of material ``stuff'' composing objects in
3D space, has become very influential in the philosophical literature
after the work of Bell and his insistence on the necessity of \emph{local
beables} and Ghirardi's notion of \emph{mass-density.} Today this program is
well known as the \emph{primitive ontology} approach\footnote{See e.g.~Allori et al.~(2008), Allori et al.~(2014), Esfeld
  (2014), Maudlin (2007).} and to some extent
we may say that Schrödinger, with his original proposal, was the first
primitive ontologist \emph{ante litteram.}

\hypertarget{polywaves-and-multi-fields}{%
\subsection{Polywaves and multi-fields}}

There is a second way through which the wave function can be linked to a
3D ontology: this is thinking (or better, re-thinking) of the wave
function per se as an object in 3D space. The first author to propose
such an interpretation has been the metaphysician Peter Forrest in his
book Quantum Metaphysics (Forrest 1988). In short, Forrest proposes to
interpret the wave function as a novel kind of field in 3D space. He
calls this interpretation ``polywave'' since in his view the wave function
gives rise to a generalized notion of a classical wave. While a
classical field assigns a precise value to each point of space, the
polywave (for an
\(n\)-particle system) assigns a precise
value to an ordered
\(n\)-tuple of points in 3D space. This can
be done by interpreting configuration space just as a mathematical
space, and the position coordinates of the wave function
\(\psi(x_1,\dots,x_N)\) as the corresponding positions of the
\emph{n} particles in 3D space.\footnote{A simple way to understand the procedure is to think of
  configuration space as the space of configuration of the particles
  in 3D space, which is exactly how configuration space is built for
  an \emph{n}-particle system in classical mechanics. The difference with
  quantum mechanics is that in the latter theory we cannot divide
  configuration space into factorized states of three coordinates.
  This does not imply however that we cannot think of those variables
  as describing positions in 3D space, rather it does imply that the
  position of the \emph{n} particles will be non-locally connected with
  each other, as we expect from quantum entanglement.} One remark is in order, however: since
in standard quantum mechanics there are no fixed/actual particles'
positions beyond the wave function, the wave function
\(\psi(x_1,\dots,x_N)\) will define an infinite set of
polywaves, each one corresponding to a possible configuration of the \emph{n}
position coordinates
\((x_1,\dots,x_N)\) of the system. This is how Forrest
defines the notion of \emph{polywave}:\footnote{Forrest (1988, p.~155).}

\begin{quote}
I posit polywaves, which are disturbances to polyadic fields. The
familiar monowaves (monadic waves) are assignments to each location of
some member of the set of possible field-values for that location.
Likewise an {[}N{]}-adic polywave is an assignment to each ordered
n-tuple of locations of a member of the set of possible field-values
for that {[}N{]}-tuple of locations. The integer {[}N{]} is just the
``number of particles''. And the possible field-values are {[}N{]}-adic
relations (Forrest 1988, p.~155).
\end{quote}

The problem with this view, in our opinion, is that the notion of
polywave is not anchored to anything in 3D space, if not an abstract
relation between \emph{n} position coordinates. The risk is to dissolve the
object in an infinite set of relations between ``empty'' position
coordinates.\footnote{This interpretation may be well suited, however, within an ontic
  structuralist approach to the wave function.} One solution to this problem is to transport Forrest's
polywave interpretation in a framework in which the \emph{n} position
coordinates correspond to the position of \emph{n} real particles in 3D
space. This can naturally be done in Bohm's theory, where an
\(n\)-particle system is mathematically
described by the \emph{n}-particle wave function
\(\psi(x_1,\dots,x_N)\) \emph{and} the position coordinates of \emph{n}
actual particles \(q=(q_1,\dots,q_N)\). Since in Bohm's theory there is only
one actual configuration of the particles at any instant of time, the
wave function will define one polywave at a time, the polywave
corresponding to the actual location of the Bohmian particles
\((q_1,\dots,q_N)\). This strategy was first formulated, and
criticized, by Belot (2012),\footnote{Belot dismisses the multi-field view (mostly) for the back
  reaction problem: while the multi-field acts on the Bohmian
  particles, the latter do not act back on the multi-field, contrary
  to the field-particle interplay that characterizes classical fields.
  This is correct, but we do not see why the multi-field should behave
  like a classical field. The main point of introducing a new field is
  to define a new entity that may (and probably will) behave
  differently from classical entities.} who gave it the name of \emph{multi-field},
and later revived and defended by Hubert \& Romano (2018). In short, the
multi-field view regards the wave function as a new kind of field in 3D
space, a field that assigns (for an
\(n\)-particle system), at any instant of
time, a precise complex value
\(c\) to the ordered \emph{n}-tuple of position
coordinates corresponding to the actual position of the Bohmian
particles:\footnote{Note that, while on the ontological level the positions of the
  Bohmian particles are fixed at any instant of time (thus we have
  only one value of the multi-field at each instant of time),
  epistemically the particles' positions are distributed according to
  the Born rule.}
\begin{equation}
\psi(q_1,\dots,q_N)=c
\end{equation}

Having complex values is not necessary for defining the multi-field: one
of the authors (D.R.) suggested, in fact, that the multi-field might be
more naturally interpreted in the original Bohm's second-order theory
(Bohm 1952), where the motion of the particles is guided by real forces
originated by real-valued fields: the amplitude
\(R(x,t)\) and phase
\(S(x,t)\) of the wave function (Romano 2021).
According to this view there are two (real-valued) multi-fields, one
corresponding to the amplitude and one corresponding to the phase of the
wave function, and the wave function is just a mathematical tool to
encode the information and coupling between these two multi-fields. A
final suggestion goes back to Schrödinger, who also regarded the idea to
split the wave function into real components as a possibility to connect
it with the ontology of the microscopic world:\footnote{Schrödinger (1926d), sect. 7: ``On the physical significance of
  the field scalar''; English translation in Schrödinger (1928, p.
  123).}

\begin{quote}
Meantime, there is no doubt a certain crudeness in the use of a
\emph{complex} wave function. If it were unavoidable \emph{in principle}, and
not merely a facilitation of the calculation, this would mean that
there are in principle \emph{two} wave functions, which must be used
\emph{together} in order to obtain information on the state of the system.
This somewhat unacceptable inference admits, I believe, of the very
much more congenial interpretation that the state of the system is
given by a real function and its time-derivative.
\end{quote}

The different realist approaches to the wave function discussed in the
paper, all inspired (directly or indirectly) by Schrödinger's proposal
in 1926,\footnote{On the plurality of ontologies proposed for Bohm's theory and
  the GRW theory, see Matarese, Hoffmann and Werner (this volume).} are listed in the following table:

\begin{table}[h]
\centering
\caption{Realist approaches to the wave function}\label{tab:realist}
\small
\begin{tabular}{@{}lll@{}}
\toprule
\textbf{Author (year)} & \textbf{Framework} & \textbf{Proposed approach}\\
\midrule
Schrödinger (1926) & Standard QM & electron charge density\\
Forrest (1988) & Standard QM & polywave\\
Ghirardi et al.\ (1995) & CSL model & mass density\\
Hubert \& Romano (2018) & Bohm's theory & multi-field\\
\bottomrule
\end{tabular}
\end{table}

\hypertarget{conclusions}{%
\section{Conclusions}}

From his earliest works on wave mechanics, Schrödinger thought of the
new quantum theory as a realist theory of the microscopic world. His
interpretation of (the absolute square of) the wave function as a
physical field in 3D space shows us that quantum theory does not force
us to renounce a precise description of matter. In fact, this served as
a concrete example for the construction of modern realist approaches to
the wave function, such as the mass-density and the multi-field
approaches. Furthermore, Schrödinger's interpretation of the electron as
electric charge density seems to be implicitly used nowadays in quantum
chemistry in formal techniques and approximations for computing the
shapes of atoms and molecules. All this is a lively part of
Schrödinger's legacy, of which this paper, on the occasion of the
centenary of quantum mechanics, is a humble tribute.

\section*{Acknowledgments}
We warmly thank Raoni Arroyo and Matteo Morganti for inviting us to contribute to this volume. The work of D. R. has been supported by the project QUANTUMARROW, ``A Quantum Origin for Time's Arrow'' (MUR/PNRR Young Researchers, SOE2024\_0000070), funded by the European Union -- NextGenerationEU and hosted by the Department of Human Sciences at the University of Verona.

\begin{refs}\bibitem Acuña, P. (forthcoming). From incompleteness to nonlocality. In R. Arroyo \& M. Morganti (Eds.), \emph{Quantum Dilemmas: Philosophical Issues in Quantum Theory after One Century}. Cham: Springer.
\bibitem Albert, D. Z. (1996). Elementary quantum metaphysics. In J. T. Cushing, A. Fine, \& S. Goldstein (Eds.), \emph{Bohmian Mechanics and Quantum Theory: An Appraisal} (pp.~277--284). Dordrecht: Kluwer.
\bibitem Allori, V., Goldstein, S., Tumulka, R., \& Zanghì, N. (2008). On the common structure of Bohmian mechanics and the Ghirardi--Rimini--Weber theory. \emph{The British Journal for the Philosophy of Science}, 59(3), 353--389.
\bibitem Allori, V., Goldstein, S., Tumulka, R., \& Zanghì, N. (2014). Predictions and primitive ontology in quantum foundations: a study of examples. \emph{The British Journal for the Philosophy of Science}, 65(2), 323-352.
\bibitem Allori, V. (forthcoming). From Einstein's statistical interpretation to the de Broglie--Bohm pilot-wave theory: Reductionism, locality, and the incompleteness of quantum mechanics. In R. Arroyo \& M. Morganti (Eds.), \emph{Quantum Dilemmas: Philosophical Issues in Quantum Theory after One Century}. Cham: Springer.
\bibitem Bacciagaluppi, G. (2025). The role of decoherence in quantum mechanics. In E. N. Zalta \& U. Nodelman (Eds.), \emph{The Stanford Encyclopedia of Philosophy} (Spring 2025 Edition). \url{https://plato.stanford.edu/archives/spr2025/entries/qm-decoherence/}
\bibitem Bassi, A., \& Ghirardi, G. (2003). Dynamical reduction models. \emph{Physics Reports}, 379(5-6), 257-426.
\bibitem Belot, G. (2012). Quantum states for primitive ontologists: A case study. \emph{European Journal for Philosophy of Science} 2(1): 67--83.
\bibitem Bohm, D. (1952). A suggested interpretation of the quantum theory in terms of ``hidden'' variables. I. \emph{Physical Review}, 85(2), 166.
\bibitem Bohm, D., Hiley, B. J., \& Kaloyerou, P. N. (1987). An ontological basis for the quantum theory. \emph{Physics Reports}, 144(6), 321-375.
\bibitem Born, M., and P. Jordan (1925). Zur Quantenmechanik. \emph{Zeitschrift für Physik} 34: 858--888.
\bibitem Born, M., W. Heisenberg, and P. Jordan (1926). Zur Quantenmechanik II. \emph{Zeitschrift für Physik} 35: 557--615.
\bibitem Callender, C. (2023). Quantum mechanics: Keeping it real? \emph{The British Journal for the Philosophy of Science} 74(4): 837--851.
\bibitem Calosi, C. (forthcoming). Varieties of wavefunction fundamentalism. In R. Arroyo \& M. Morganti (Eds.), \emph{Quantum Dilemmas: Philosophical Issues in Quantum Theory after One Century}. Cham: Springer.
\bibitem Crull, E. M. (2013). Exploring philosophical implications of quantum decoherence. \emph{Philosophy Compass} 8(9): 875--885.
\bibitem De Broglie, L. (1924). \emph{Recherches sur la théorie des quanta}. Paris: Sorbonne.
\bibitem Duncan, A., and M. Janssen (2023). \emph{Constructing Quantum Mechanics. Volume Two: The Arch, 1923--1927}. Oxford: Oxford University Press.
\bibitem Einstein, A. (1924). Quantentheorie des einatomigen idealen Gases I. \emph{Sitzungsberichte der Preußischen Akademie der Wissenschaften}: 261--267.
\bibitem Einstein, A. (1925). Quantentheorie des einatomigen idealen Gases II. \emph{Sitzungsberichte der Preußischen Akademie der Wissenschaften}: 3--14.
\bibitem Esfeld, M. (2014). The primitive ontology of quantum physics: guidelines for an assessment of the proposals. \emph{Studies in History and Philosophy of Science Part B: Studies in History and Philosophy of Modern Physics}, 47, 99-106.
\bibitem Forrest, P. (1988). \emph{Quantum Metaphysics}. New York: Blackwell.
\bibitem Ghirardi, G. C., Rimini, A., \& Weber, T. (1986). Unified dynamics for microscopic and macroscopic systems. \emph{Physical Review D}, 34(2), 470.
\bibitem Ghirardi, G. C., Pearle, P., \& Rimini, A. (1990). Markov processes in Hilbert space and continuous spontaneous localization of systems of identical particles. \emph{Physical Review A}, 42(1), 78--89.
\bibitem Ghirardi, G. C., Grassi, R., \& Benatti, F. (1995). Describing the macroscopic world: closing the circle within the dynamical reduction program. \emph{Foundations of Physics}, 25(1), 5-38.
\bibitem Hanle, P. A. (1977). The coming of age of Erwin Schrödinger: His quantum statistics of ideal gases. \emph{Archive for History of Exact Sciences} 17: 165--192.
\bibitem Heisenberg, W. (1925). Über quantentheoretische Umdeutung kinematischer und mechanischer Beziehungen. \emph{Zeitschrift für Physik} 33: 879--893.
\bibitem Hubert, M., and D. Romano (2018). The Wave Function as a Multi-field. \emph{European Journal for Philosophy of Science} 8(3): 521--537.
\bibitem Joas, C., and C. Lehner (2009). The classical roots of wave mechanics: Schrödinger's transformations of the optical-mechanical analogy. \emph{Studies in History and Philosophy of Science Part B: Studies in History and Philosophy of Modern Physics} 40: 338--351.
\bibitem Lombardi, O., \& Fortin, S. (forthcoming). Schrödinger on electron density: Current relevance in quantum chemistry. In R. Arroyo \& M. Morganti (Eds.), \emph{Quantum Dilemmas: Philosophical Issues in Quantum Theory after One Century}. Cham: Springer.
\bibitem Matarese, V., Hoffmann, V., \& Werner, J. (forthcoming). Nomic indeterminacy in quantum mechanics: A centennial problem. In R. Arroyo \& M. Morganti (Eds.), \emph{Quantum Dilemmas: Philosophical Issues in Quantum Theory after One Century}. Cham: Springer.
\bibitem Maudlin, T. W. (2007). Completeness, supervenience and ontology. Journal of Physics A: Mathematical and Theoretical, 40(12), 3151-3171.
\bibitem Mehra, J., and H. Rechenberg (1987a). \emph{The Historical Development of Quantum Theory}, Vol. 5, Part 1: \emph{Erwin Schrödinger and the Rise of Wave Mechanics: Schrödinger in Vienna and Zürich, 1887--1925}. Heidelberg: Springer.
\bibitem Mehra, J., and H. Rechenberg (1987b). \emph{The Historical Development of Quantum Theory}, Vol. 5, Part 2: \emph{Erwin Schrödinger and the Rise of Wave Mechanics: The Creation of Wave Mechanics: Early Response and Applications, 1925--1926}. Heidelberg: Springer.
\bibitem Ney, A. (2021). \emph{The World in the Wave Function: A Metaphysics for Quantum Physics}. New York: Oxford University Press.
\bibitem Norsen, T. (2017). Foundations of Quantum Mechanics. Springer.
\bibitem Pearle, P. (1989). Combining stochastic dynamical state-vector reduction with spontaneous localization. \emph{Physical Review A}, 39(5), 2277--2289.
\bibitem Przibram, K. (1967). \emph{Letters on Wave Mechanics: Schrödinger, Planck, Einstein, Lorentz}. New York: Philosophical Library.
\bibitem Romano, D. (2021). Multi-field and Bohm's theory. Synthese, 198(11), 10587-10609.
\bibitem Romano, D. (2022). The unreasonable effectiveness of decoherence. In V. Allori (Ed.), Quantum Mechanics and Fundamentality: Naturalizing Quantum Theory between Scientific Realism and Ontological Indeterminacy (pp.~3-18). Cham: Springer International Publishing.
\bibitem Romano, D. (2023). A decoherence-based approach to the classical limit in Bohm's theory. Foundations of Physics, 53(2), 41.
\bibitem Rusi El Hassani, A., \& López, C. (forthcoming). The case against space realism: Heisenberg and algebraic quantum theory. In R. Arroyo \& M. Morganti (Eds.), \emph{Quantum Dilemmas: Philosophical Issues in Quantum Theory after One Century}. Cham: Springer.
\bibitem Schlosshauer, M. (2007). Decoherence and the quantum-to-classical transition. Springer: Berlin-Heidelberg.
\bibitem Schlosshauer, M. (2019). Quantum decoherence. Physics Reports, 831, 1-57.
\bibitem Schrödinger, Erwin (1926a) `Quantisierung als Eigenwertproblem. Erste Mitteilung', \emph{Annalen der Physik}, 79: 361--76.
\bibitem Schrödinger, Erwin (1926b) `Quantisierung als Eigenwertproblem. Zweite Mitteilung', \emph{Annalen der Physik}, 79: 489--527.
\bibitem Schrödinger, Erwin (1926c) `Quantisierung als Eigenwertproblem. Dritte Mitteilung', \emph{Annalen der Physik}, 80: 437--90.
\bibitem Schrödinger, Erwin (1926d) `Quantisierung als Eigenwertproblem. Vierte Mitteilung', \emph{Annalen der Physik}, 81: 109--39.
\bibitem Schrödinger, E. (1928). \emph{Collected Papers on Wave Mechanics}. London: Blackie \& Son.
\bibitem Scott, W. T. (1967). \emph{Erwin Schrödinger: An Introduction to His Writings}. Amherst: University of Massachusetts Press.
\bibitem Sebens, C. T. (2021). Electron charge density: A clue from quantum chemistry for quantum foundations. \emph{Foundations of Physics} 51(4): 75.
\bibitem Simon, D. (ed.) (2006). \emph{Albert Einstein: Akademie-Vorträge. Sitzungsberichte der Preußischen Akademie der Wissenschaften, 1914--1932}. Weinheim: Wiley-VCH.
\bibitem Wallace, D. (2012). The Emergent Multiverse: Quantum theory according to the Everett interpretation. Oxford University Press, Oxford.
\bibitem Wessels, L. (1979). Schrödinger's route to wave mechanics. \emph{Studies in History and Philosophy of Science Part A} 10: 311--340.
\bibitem Zeh, H. D. (1993). There are no quantum jumps, nor are there particles!. \emph{Physics Letters A}, 172(4), 189-192.
\bibitem Zurek, W. H. (1991). Decoherence and the transition from quantum to classical. \emph{Physics Today}, 44(10), 36-44.
\bibitem Zurek, W. H., Habib, S., \& Paz, J. P. (1993). Coherent states via decoherence. \emph{Physical Review Letters}, 70(9), 1187.\end{refs}
\end{document}